\documentclass[aps,prd,twocolumn,amsmath,amssymb,showpacs,nofootinbib]{revtex4}

\usepackage{graphicx,slashed,cancel,bbm,hyperref,framed,xcolor,mathrsfs}
\usepackage[all]{xy}

\newcommand{\be}{\begin{equation}}
\newcommand{\ee}{\end{equation}}

\newcommand{\D}{\mathrm{d}}
\newcommand{\E}{\mathrm{e}}
\newcommand{\I}{\mathrm{i}}

\newcommand{\Tr}{\mathrm{Tr}}
\newcommand{\Lag}{\mathcal{L}}

\newcommand{\V}{{\mathcal{V}}}

\newcommand{\W}{{\mathcal{W}}}

\newcommand{\Z}{{\mathcal{Z}}}

\newcommand{\g}{\mathfrak{g}}
\newcommand{\fg}{\bar{\mathfrak{g}}}
\newcommand{\old}{{\varepsilon_\mathrm{old}}}
\newcommand{\new}{{\varepsilon_\mathrm{new}}}
\newcommand{\sfrac}[2]{{\textstyle\frac{#1}{#2}}}

\newcommand{\four}{\circ}
\newcommand{\five}{\bullet}

\newcommand{\opt}{\mathfrak}

\begin{document}
\title{Schwinger's proper time and worldline holographic renormalisation}
\author{Dennis D. Dietrich$^a$}
\author{Adrian Koenigstein$^{ab}$}
\affiliation{$^a$Institut f\"ur Theoretische Physik, Goethe-Universit\"at, Frankfurt am Main, Germany}
\affiliation{$^b$Frankfurt Institute for Advanced Studies, Frankfurt am Main, Germany}
\begin{abstract}
Worldline holography states that within the framework of the worldline approach to quantum field theory, sources of a quantum field theory over Mink$_4$ naturally form a field theory over AdS$_5$ {\sl to all orders} in the elementary fields and in the sources of arbitrary spin. (Such correspondences are also available for other pairs of spacetimes, not only Mink$_4\leftrightarrow\mathrm{AdS}_5$.) Schwinger's proper time of the worldline formalism is automatically grouped with the physical four spacetime dimensions into an AdS$_5$ geometry. We show that the worldline holographic effective action in general and the proper-time profiles of the sources in particular solve a renormalisation group equation and, reversely, can be defined as solution to the latter. This fact also ensures regulator independence.
\end{abstract}
\maketitle
\section{Introduction}
Strong interactions are behind a wealth of phenomena, but oftentimes beyond our computational abilities. Holographic approaches promise analytic insight and have been applied, for example, to quantum chromodynamics (QCD) \cite{Erlich:2005qh,Karch:2006pv,Polchinski:2000uf}, extensions of the Standard Model \cite{Hong:2006si,Dietrich:2008ni}, condensed-matter physics \cite{Sachdev:2011wg}, and the Schwinger effect \cite{Sato:2013dwa,Gorsky:2001up,Dietrich:2014ala}. Holography takes off from the conjectured AdS/CFT correspondence \cite{'tHooft:1973jz,Maldacena:1997re,Witten:1998qj}.  All examples of this correspondence found since, however, hold for theories with a set of symmetries that are not found in nature and which posses a particle content different from QCD. As a result, in practice, deformed bottom-up AdS/QCD descriptions are considered, which describe the QCD hadron spectrum rather well \cite{Karch:2006pv,Da Rold:2005zs}. Still, they lack a derivation from first principles. As a consequence, it is very important to comprehend under which circumstances and for which reasons these models represent an acceptable approximation and which features are robust. For some approaches to these questions see \cite{deTeramond:2008ht,deTeramond:2013it,Cata:2006ak}.

In this context, \cite{Dietrich:2014ala,Dietrich:2013kza,Dietrich:2013vla,Dietrich:2015oba} demonstrated, how a quantum field theory over Mink$_4$ naturally turns into a field theory for its sources over AdS$_5$ in the framework of the worldline formalism \cite{Feynman:1950ir,Strassler:1992zr} for quantum field theory. Schwinger's proper time naturally becomes the fifth dimension of an AdS$_5$ geometry. This result also extends to different pairs of spacetimes including the non-relativistic case. \cite{Dietrich:2015oba} showed that such an AdS$_5$ formulation arises {\sl to all orders} in the elementary fields---matter and gauge. Schwinger's proper time represents a length scale (inverse energy scale), which is also the interpretation of the extra dimension in holography \cite{Erlich:2005qh,Karch:2006pv,Polchinski:2000uf,Maldacena:1997re,Witten:1998qj}. Handling ultraviolet (UV) divergences of a theory necessitates regularisation, which in the worldline formalism is customarily taken care of by proper-time regularisation, the introduction of a positive lower bound on the proper-time. The proper-time regularisation corresponds to the UV-brane regularisation in holography \cite{Erlich:2005qh,Karch:2006pv,Polchinski:2000uf,Maldacena:1997re,Witten:1998qj}. The worldline holographic framework ensures the independence of physical predictions from the unphysical value of this cutoff parameter, which identifies worldline holography as a renormalisation group flow. 

This central statement of the present paper is treated in Sect.~\ref{sec:fifth}. Before that, in Sect.~\ref{sec:wlholo}, building on \cite{Dietrich:2015oba}, we reiterate on how exactly a quantum field theory over Mink$_4$ reorganises into a field theory for its sources over AdS$_5$ {\sl to all orders} in the elementary fields and the sources and with Schwinger's proper time of the worldline formalism as fifth dimension. After that, in Sect.~\ref{sec:free}, we analyse some examples in the free case. First, in Sect.~\ref{sec:qed}, we study the holographic renormalisation of quantum electrodynamics (QED). Subsequently, in Sect.~\ref{sec:hs} we turn to general higher-spin sources. Particularly, we show that the AdS$_5$ geometry is self-consistent. The penultimate section \ref{sec:sum} provides a short summary. We conclude with a Discussion and Outlook in Sect.~\ref{sec:out}.

\section{Worldline holography\label{sec:wlholo}}

In order to display the essence of the program we begin with one massless scalar\footnote{Nothing obstructs the treatment of elementary matter with spin, but, for the sake of simplicity and clarity, we here limit ourselves to spinless elementary matter. Results for fermionic matter are presented elsewhere \cite{Master,Fermions}.} flavour and a vector source $V$ combined with the gauge field $G$ in the `covariant derivative' $\mathbbm{D}=\partial-\I\mathbbm{V}$, where $\mathbbm{V}=G+V$. Thus, to all orders, the generating functional for vector correlators reads 
\begin{align}
Z=\langle\E^w\rangle=\int[\D G]\E^{w-\frac{i}{4e^2}\int\D^4x\, G_{\mu\nu}^2},
\label{eq:genfun}
\end{align}
where 
\begin{align}
w
=
-\frac{1}{2}\Tr\ln\mathbbm{D}^2.
\label{eq:scalar}
\end{align}
In the worldline formalism\footnote{Particle-wave duality. The worldline representation \eqref{eq:wl} of the functional determinant \eqref{eq:scalar} can actually be seen as the particle dual of the determinant's wave(-function or field) representation as Feynman functional integral
$$
w=\ln\int[\D\phi][\D\phi^\dagger]\,\E^{\I\int\D^4x\,\phi^\dagger D^2\phi}.
$$} \cite{Feynman:1950ir,Strassler:1992zr} after a Wick rotation, $w$ is given by \cite{Dietrich:2014ala,Dietrich:2013kza,Dietrich:2013vla}
\begin{align}
w
=
&\int \D^4x_0\int_{\varepsilon>0}^\infty\frac{\D T}{2T^3}
\,\Lag\equiv\iint_\varepsilon^\infty\D^5x\,\sqrt{g}
\,\Lag ,
\label{eq:wl}
\\
\Lag
=
&\frac{2\mathcal{N}}{(4\pi)^2}
\int_\mathrm{P}[\D y]\;\E^{-\int_0^T\D\tau[\frac{\dot y^2}{4}+\I \dot y \cdot\mathbbm V(x_0+y)]} ,
\label{eq:lag}
\end{align}
with the five-dimensional metric $g$
\be
\D s^2\overset{g}{=}+\frac{\D T^2}{4T^2}+\frac{\D x_0\cdot\D x_0}{T}
\label{eq:patch}
\ee
and the square root of the absolute of its determinant $\sqrt{g}$.
`$\cdot$' represents the contraction with the flat four-dimensional metric $\eta_{\mu\nu}$, Wick rotated from mostly plus to all plus, which, simultaneously, turns \eqref{eq:patch} from an AdS$_{4,1}$ to an AdS$_{5,0}$ line element.
The isometries of the five-dimensional AdS space are the symmetries of the conformal group of the corresponding four-dimensional flat space.  
$T$ stands for Schwinger's proper time. A factor of $T^{-1}$ in the volume element arises from exponentiating the logarithm in \eqref{eq:scalar}, another factor of $T^{-2}$ from taking the trace. 
The Lagrangian density $\mathcal{L}$ is made up of a the path integral over all closed paths over the proper-time interval $[0;T]$, i.e., with $y(0)=y(T)$.
The $\D^4x_0$ integral translates these paths to every position in space, $x\equiv y+x_0$. (The translations are the zero modes of the kinetic operator $\partial_\tau^2$. $\dot y\equiv\partial_\tau y$. Splitting them off from the rest of the path integral also makes momentum conservation manifest.)\footnote{For details and more intermediate steps please see \cite{Dietrich:2013kza,Dietrich:2015oba,Strassler:1992zr}.\label{foot3}} 
Inspection of the free part of the worldline action $\int_0^T\D\tau\frac{(\partial_\tau y)^2}{4}=\int_0^1\D\hat\tau\frac{(\partial_{\hat\tau} y)^2}{4T}$, where $\hat\tau=\tau/T$, shows that small values of $T$ correspond to short relative distances, i.e., to the UV regime. Thus, the proper-time regularisation $T\ge\varepsilon>0$ (introduced when exponentiating the logarithm$^{\ref{foot3}}$) is a UV regularisation and corresponds to the UV-brane regularisation in holography \cite{Karch:2006pv,Erlich:2005qh,Polchinski:2000uf}. 

\subsection{Volume elements}

Thus, in the worldline formalism, $w$ falls readily into the form of an action \eqref{eq:wl} over AdS$_5$. $e^w$, however, contains all powers of $w$. Building on the discussion in \cite{Dietrich:2015oba}, for the $n$-th power,
\begin{align}
w^n
=&
\prod_{j=1}^n
\int\D^4x_j\int_{\varepsilon}^\infty\frac{\D T_j}{2T_j^3}
\,\Lag(x_j,T_j).
\end{align}
The free part again only depends on the relative positions and we split off a common absolute coordinate $x_0=\opt{x_0}(\{x_j\})$, which can be any linear combination of the $x_j$, e.g., the centre of mass $\frac{1}{n}\sum_{j=1}^n x_j$. There remain integrations over $n-1$ four-dimensional relative coordinates, $\D^{4(n-1)}\Delta$,
\begin{align}\nonumber
&\int \prod_{j=1}^n\D^4x_j
=\\=
&\int \prod_{j=1}^n\D^4x_j\int\D^4x_0\delta^{(4)}[x_0-\opt{x_0}(\{x_k\})]
=\\=
&\int\D^4x_0\int\Big(\prod_{j=1}^n\D^4x_j\Big)\delta^{(4)}[x_0-\opt{x_0}(\{x_k\})]
=\\=
&\int\D^4x_0\int\D^{4(n-1)}\Delta.
\end{align}

Analogously, we introduce an overall proper time $T=\opt{T}(\{T_j\})$ and proper-time fractions $t_j=T_j/T$ using
\be
1=\int\D T\,\delta[T-\opt{T}(\{T_j\})]\prod_{j=1}^n\Big[\int\D t_j\,\delta\Big(t_j-\frac{T_j}{T}\Big)\Big].
\label{eq:choco}
\ee
Also here, there exists a continuum of choices for the overall proper time $T$, all of which allow us to come to the same conclusion below. Introducing additional dimensionfull scales would be artificial. In their absence, on dimensional grounds always $\opt{T}(\{T_j\})=T\times\opt{T}(\{t_j\})$. A definition invariant under the pairwise exchange of the $T_j$ makes the corresponding symmetry of $w^n$ manifest from the start. The arguably simplest choice with these characteristics would be $\opt{T}=\sum_{j=1}^nT_j$. In any case, physics is invariant under any invertible change of variables. \eqref{eq:choco} implies
\begin{align}\nonumber
&\prod_{j=1}^n\int_{\varepsilon}^\infty\frac{\D T_j}{2T_j^3}
=\\=
&\int\D T\prod_{j=1}^n\Big[\int_{\varepsilon}^\infty\frac{\D T_j}{2T_j^3}\int\D t_j\,\delta\Big(t_j-\frac{T_j}{T}\Big)\Big]\times\\\nonumber&\hskip 5cm\times
\delta[T-\opt{T}(\{T_l\})]
=\\=
&\int\frac{\D T}{2T^3}T^{-2(n-1)}\int_{\frac{\varepsilon}{T}}^\infty\Big(\prod_{j=1}^n\frac{\D t_j}{2t_j^3}\Big)\:2\delta[1-\opt{T}(\{t_l\})].
\end{align}
Putting everything together,
\begin{align}
w^n
=&
\int\D^4x_0\int\frac{\D T}{2T^3}\int\frac{\D^{4(n-1)}\Delta}{T^{2(n-1)}}
\int_{\frac{\varepsilon}{T}}^\infty\Big[\prod_{j=1}^n\frac{\D t_j}{2t_j^3}
\times\\\nonumber&\hskip0.75cm\times
\,\Lag(x_0+x_j-x_0,Tt_j)\Big]
2\delta[1-\opt{T}(\{t_l\})]
=\\=&
\int\D^4x_0\int\frac{\D T}{2T^3}\int\D^{4(n-1)}\hat\Delta
\int_{\frac{\varepsilon}{T}}^\infty\Big[\prod_{j=1}^n\frac{\D t_j}{2t_j^3}
\times\\\nonumber&\hskip0.75cm\times
\,\Lag(x_0+\widehat{x_j-x_0}\sqrt{T},Tt_j)\Big]
2\delta[1-\opt{T}(\{t_l\})],
\end{align}
where $x_j-x_0$ in the argument of $\Lag_j$ depends only on the relative coordinates $\Delta$ and not the absolute coordinate $x_0$. In the last step, we introduced dimensionless relative coordinates $\hat\Delta=\Delta/\sqrt{T}$. 
This demonstrates that every power $w^n$ takes the form of a Lagrangian density integrated over AdS$_5$.

\subsection{Contractions}

It remains to be shown that all spacetime indices are contracted with AdS metrics. Expressing the vector $\mathbbm{V}$ in $\Lag(x_j,T_j)$ in terms of dimensionless variables, $\widehat{x_j-x_0}$ as well as $\hat y_j=y_j/\sqrt{T_j}=y_j/\sqrt{Tt_j}$, and with the help of a translation operator, 
\begin{align}
\mathbbm{V}(y_j+x_j)
&=
\mathbbm{V}(y_j+x_j-x_0+x_0)
=\\&=
\mathbbm{V}[\sqrt{T}(\hat y_j\sqrt{t_j}+\widehat{x_j-x_0})+x_0]
=\\&=
\E^{\sqrt{T}(\hat y_j\sqrt{t_j}+\widehat{x_j-x_0})\cdot\partial_{x_0}}\mathbbm{V}(x_0),
\end{align}
we obtain from \eqref{eq:lag},
\begin{align}\nonumber
&\Lag(x_j,T_j)
=\\={}&
\frac{2\mathcal{N}}{(4\pi)^2}
\int_\mathrm{P}[\D y_j]\;\E^{-\int_0^T\D\tau[\frac{\dot y^2}{4}+\I \dot y \cdot\mathbbm V(y_j+x_j)]}
=\\={}&
\frac{2\hat{\mathcal{N}}}{(4\pi)^2}
\int_\mathrm{P}[\D\hat y_j]
\\\nonumber
&\E^{-\int_0^1\D\hat\tau_j[\frac{(\partial_{\hat\tau_j}\hat y_j)^2}{4}+\I\sqrt{t_jT}(\partial_{\hat\tau_j}\hat y_j)\cdot\E^{\sqrt{T}(\sqrt{t_j}\hat y_j+\widehat{x_j-x_0})\cdot \partial_{x_0}}\mathbbm V(x_0)]},
\end{align}
where we also use the dimensionless integration variable $\hat\tau_j=\tau_j/T_j=\tau_j/(Tt_j)$. This expression shows that every gradient $\partial_{x_0}$ and every field $\mathbbm V$, i.e., every four-dimensional spacetime index, is accompanied by one power of $\sqrt{T}$. 

The same holds still after integrating out the gauge field $G$. In order to see this, we split the Wilson line for $\mathbbm{V}$ into one for the sources $V$ and one for the gauge fields $G$,
\begin{align}
\E^{\I\oint\D x\cdot\mathbbm{V}}
=
\E^{\I\oint\D x\cdot V}
\E^{\I\oint\D x\cdot G}.
\end{align}
This is possible, because the sources $V$ are gauge singlets and the gauge fields $G$ are flavour singlets and commute as a consequence. Then, with the definition of the gauge-field average from \eqref{eq:genfun},
\begin{align}
\nonumber
&\Big\langle\prod_{j=1}^n\Lag_j\Big\rangle
=
\bigg(\frac{2\hat{\mathcal{N}}}{(4\pi)^2}\bigg)^n
\prod_{j=1}^n\int_\mathrm{P}[\D\hat y_j]
\nonumber\\&\nonumber
\E^{-\int_0^1\D\hat\tau_j [\frac{(\partial_{\hat\tau_j}\hat y_j)^2}{4}+\I\sqrt{t_jT}(\partial_{\hat\tau_j}\hat y_j)\cdot\E^{\sqrt{T}(\sqrt{t_j}\hat y_j+\widehat{x_j-x_0})\cdot \partial_{x_0}}V(x_0)]}
\\&
\times
\Big\langle
\prod_{l=1}^n\E^{\I\oint\D y_l\cdot G(x_l+y_l)}
\Big\rangle.
\label{eq:aint}
\end{align}
In the factor on the last line, $G$ is integrated out. Said factor is invariant under reparametrisations of the Wilson line
as well as of four-dimensional translations. As a consequence, it is independent from the value of $T$ as well as from $x_0$. Hence, it only depends on the four-dimensional relative coordinates. Additionally, the factor is a scalar and as such can only depend on the combinations ($\forall~l,j$)
\begin{align}\nonumber
&\eta^{\mu\nu}(y_j+x_j-x_0)_\mu(y_l+x_l-x_0)_\nu
=\\=
{}&T\eta^{\mu\nu}(\sqrt{t_j}\hat y_j+\widehat{x_j-x_0})_\mu(\sqrt{t_l}\hat y_l+\widehat{x_l-x_0})_\nu.
\end{align}

~\\

Taking stock, the powers of $\sqrt{T}$ stay (only) with every $\partial_{x_0}$ and $V$. 
Thus, after the $[\D\hat y]$ integrations, the result can only contain the combinations $T\eta^{\mu\nu}V_\mu V_\nu$, $T\eta^{\mu\nu}V_\mu \partial_\nu$, and $T\eta^{\mu\nu}\partial_\mu \partial_\nu$, which combine into $g^{\mu\nu}V_\mu V_\nu$, $g^{\mu\nu}V_\mu \partial_\nu$, and $g^{\mu\nu}\partial_\mu \partial_\nu$, respectively\footnote{The derivatives always act on some source $V$.}. Consequently, \eqref{eq:genfun} can be expressed as an action over AdS$_5$ for its sources to all orders and to all orders in the elementary fields. 

\section{The fifth dimension\label{sec:fifth}}

Previously, \cite{Dietrich:2014ala,Dietrich:2013kza,Dietrich:2015oba} identified the fifth-components of gradients and fields as arising from an induced Wilson (gradient) flow and determined by a variational principle. Here we show that the identical result is obtained by imposing the independence of the effective action from the ultraviolet proper-time regulator $\varepsilon$. Ultimately, this amounts to a Wilson-Polchinsky renormalisation condition.

Taking into account the above findings, organised in an expansion with respect to gradients and sources, \eqref{eq:genfun} can be expressed as
\begin{align}
\label{eq:genfuncser}
Z_\varepsilon
=
\int\D^4x_0\int_\varepsilon^\infty&\D T\sqrt{g}\sum_{n_\partial,n_V} \#_{n_\partial,n_V}
\times\\&~~~\times\nonumber
 (g^{\four\four})^\frac{n_\partial+n_V}{2}(\partial_\four)^{n_\partial}[V_\four(x_0)]^{n_V}.
\end{align}
There are only contributions from $n_\partial+n_V$ even.
The $\#_{n_\partial,n_V}$ are dimensionless numerical coefficients obtained after integrating out all proper-time fractions $t_j$ and $\hat\tau_j$ as well as dimensionless relative coordinates $\hat\Delta$. The indices `$\four$' indicate that the contractions with the five-dimensional (inverse) metric $g$ are only executed in four dimensions. (The addends in the previous expression symbolise the occurring combinations of contractions. Also, not all the derivatives act on all the sources.)

This expression contains the proper-time regulator $\varepsilon>0$, whose value possesses a priori no physical meaning. As a consequence, $Z_\varepsilon$ must not depend on the value of $\varepsilon$, i.e., $Z_\old\overset{!}{=}Z_\new$ for $\old\neq\new$. In order to study the consequences of this requirement let us try to undo the change $\old\rightarrow\new$ in
\begin{align}
Z_\new
=
\int\D^4x_0\int_\new^\infty&\frac{\D T}{2T^3}\sum_{n_\partial,n_V}\#_{n_\partial,n_V}
\times\\\times\nonumber
& (T\eta^{\four\four})^\frac{n_\partial+n_V}{2}(\partial_\four)^{n_\partial}[V_\four(x_0)]^{n_V}.
\end{align}
(This is to be accomplished for all configurations $V$. Therefore, the independence must be enforced order by order, i.e., $\forall~n_\partial,n_V$ separately.) To this end, we need to change the integration bound without changing the integrand. 

A global rescaling of the integration variables,
\begin{align}
T\rightarrow c_T T~~~~~\mathrm{as~well~as}~~~~~x_0\rightarrow c_x x_0
\end{align}
and consequently
\begin{align}
\partial_{x_0}\rightarrow c_x^{-1}\partial_{x_0}
\end{align}
leads to
\begin{align}
Z_\new=&\int\D^4x_0\int_{c_T\new}^\infty\frac{\D T}{2T^3}\sum_{n_\partial,n_V}\#_{n_\partial,n_V}
\times\\\times\nonumber
&  c_x^{4-n_\partial}c_T^{\frac{n_\partial+n_V}{2}-2}(T\eta^{\four\four})^\frac{n_\partial+n_V}{2}(\partial_\four)^{n_\partial}[V_\four(x_0)]^{n_V}.
\end{align}
Restoring the original integration bound requires $c_T=\old/\new$. Independence from $n_\partial$ necessitates $c_x=c_T^{1/2}$, which also takes care of the $n_\partial$-independent factors from the integration measure. Independence from $n_V$ can only be obtained by also rescaling $V\rightarrow c_x^{-1}V$, i.e., like the partial derivative\footnote{\label{foot1}
The interaction part in \eqref{eq:lag}, $\E^{-\I\oint\D y\cdot V}$, being a Wilson loop, is manifestly locally invariant under the transformation $V^\mu\rightarrow\Omega[V^\mu+\I\Omega^\dagger(\partial^\mu\Omega)]\Omega^\dagger$, which entails hidden local symmetry \cite{Bando:1984ej}. Therefore, an alternative expression only using covariant derivatives is also available \cite{Shifman:1980ui,Schmidt:1993rk},
$$
Z=\iint_\varepsilon^\infty\D^5x\sqrt{g}\sum_n \#_n (g^{\four\four})^n(D_\four)^{2n},
$$
where $D=\partial-\I V$.
This corroborates why $V$ must scale like the partial derivative.
Moreover, the proper-time regularisation keeps this symmetry manifest, at variance with a momentum cutoff. 
}.
Thus, regulator independence of $Z_\varepsilon$ can be achieved,
\begin{align}
Z_\new=&\int\D^4x_0\int_{\old}^\infty\frac{\D T}{2T^3}\sum_{n_\partial,n_V}\#_{n_\partial,n_V}
\times\\&\hskip 1.4cm\times\nonumber
(T\eta^{\four\four})^\frac{n_\partial+n_V}{2}(\partial_\four)^{n_\partial}[V_\four(x_0;\old)]^{n_V},
\end{align}
but the source $V$ must rescale as well and thus depend on the value of the regulator, which is a fifth-dimensional quantity;
\begin{align}
\old V(x_0;\old)=\new V(x_0;\new).
\end{align}

\subsection{Using AdS isometries\label{sec:ads}}

Given that we had already recognised that $Z$ takes the form of an action over AdS$_5$ \cite{Dietrich:2014ala,Dietrich:2013kza,Dietrich:2013vla,Dietrich:2015oba} and that the isometries of AdS$_5$ coincide with the conformal group over Mink$_4$, which includes the invariance under scale transformations, the above approach is rather pedestrian. Introducing the aforementioned missing ingredients of fifth-dimensional gradients and components into \eqref{eq:genfuncser} would complete the field theory over AdS$_5$, which then features all the isometries of that spacetime. 

In the four-dimensional theory, however, there were no fifth-dimensional polarisations. In order to be allowed to omit them, $\V_T=0$ must be an admissible gauge condition. That means, the extension to five dimensions must feature five-dimensional local invariance under the flavor group. Due to the previously present four-dimensional invariance, this is achieved by\footnote{This is even more clear-cut in the representation given in footnote \ref{foot1}, where one would replace all flavour covariant derivatives by flavour and generally covariant derivatives.} 
\begin{align}
\label{eq:calz}
\Z
=
\iint_\varepsilon^\infty\D^5x\sqrt{g}&\sum_{n_\partial,n_V}\#_{n_\partial,n_V}
\times\\&\times
 (g^{\five\five})^\frac{n_\partial+n_V}{2}(\nabla_\five)^{n_\partial}[\V_\five(x_0,T)]^{n_V},
\nonumber
\end{align}
where the indices `$\five$' indicate the full five-dimensional contraction, and 
$\nabla$ represents the AdS covariant derivative. 
\eqref{eq:calz} does not depend on the value of $\varepsilon$ if $\V(x_0,T)$ transforms like a five-dimensional vector. 
(We would like to point out that we did {\sl not} include an explicit dependence on $\varepsilon$ in $\V$.)
If we impose the $\V_T=0$ gauge already at the level of the action, the desired scale invariance is still manifest, as scale transformations do not mix tensor components, while the special conformal transformations do. 
The full invariance, however, is still present modulo a subsequent local flavour transformation.

As it stands, $\Z$ is merely a functional of arbitrary source configurations $\V$. The true meaning of an action for its field theory is through the configuration (or configurations) it distinguishes as saddle points, $\breve\V$. 
The boundary condition
\begin{align}
\breve\V_\mu(x_0,T=\varepsilon)=V_\mu(x_0)
\label{eq:inicond}
\end{align}
communicates the four-dimensional polarisations to the five-dimensional field $\V$ and gives it the same normalisation like $V$, i.e., as the source for {\sl once} the vector current.
It is also consistent with the previous findings in the context of worldline holography \cite{Dietrich:2014ala,Dietrich:2013kza,Dietrich:2015oba} that the worldline expressions satisfy a Wilson (gradient) flow equation with this boundary condition. 

Accordingly, the cutoff independent effective action is given by $\Z$ evaluated on the saddle point configuration in $\breve\V_T=0$ gauge,
\begin{align}
\breve\Z
=
\iint_\varepsilon^\infty\D^5x\sqrt{g}&\sum_{n_\partial,n_V}\#_{n_\partial,n_V}
\times\\\nonumber&\times
 (g^{\five\five})^\frac{n_\partial+n_V}{2}(\nabla_\five)^{n_\partial}[\breve\V_\four(x_0,T)]^{n_V}.
\end{align} 

\eqref{eq:inicond} also puts the bare source configuration at the ultraviolet end of the fifth dimension (in the aforediscussed sense that small values of $T$ correspond to short four-dimensional distances). This fact together with the requirement that the effective action do not depend on the value of the unphysical UV regulator $\varepsilon$, which can also be expressed in differential form, 
\be
\varepsilon\partial_\varepsilon \ln Z_\varepsilon\overset{!}{=}0,
\label{eq:renorm}
\ee
makes this a Wilson-Polchinsky renormalisation condition \cite{Wilson:1971bg}.

Finally, this is also the boundary condition imposed in holography \cite{Karch:2006pv,Erlich:2005qh,Polchinski:2000uf,Maldacena:1997re,Witten:1998qj}: the effective action for the four-dimensional side of the holographic duality is described by the five-dimensional action evaluated on its saddle point. Worldline holography identifies Schwinger's proper time with the fifth dimension \cite{Dietrich:2014ala,Dietrich:2013kza,Dietrich:2013vla,Dietrich:2015oba} and the fifth-dimensional profile of the sources as solution to the renormalisation group equation \eqref{eq:renorm}.

\subsection{Free case\label{sec:free}}

In order to flesh out the above presented formalism, we analyse here the free case. It is obtained from \eqref{eq:genfun} by switching off the coupling to the gauge bosons $G$. Consequently, here it is sufficient to study $w$ with $\mathbbm{V}=V$.\footnote{Based on the observation that at low energies the contributions with the lowest number of exchanged gauge bosons dominate \cite{Okubo:1963fa,Shifman:1978bx,De Rujula:1975ge,Dietrich:2012un} these are the kinematically dominant diagrams in that regime.}

\subsubsection{Holographic 1-loop renormalisation of scalar QED\label{sec:qed}}

When identifying the vector source $V$ with a (background) gauge field, $w$ is the QED 1-loop effective action. The (logarithmically) UV divergent piece is given by
\begin{align}
Z_\varepsilon=\#_{2,2}\iint_\varepsilon^\infty\D^5x\sqrt{g}g^{\mu\kappa}g^{\nu\lambda}V_{\mu\nu}V_{\kappa\lambda},
\label{eq:qedz}
\end{align}
where $V_{\mu\nu}$ stands for the (here Abelian) field-strength tensor. (The logarithmic divergence arises in the $\D T$ integration, where there is a factor of $T^{-3}$ from the volume element and two factors of $T$, one from each metric $g^{\mu\nu}$.) For the contribution from $N_f\times N_c$ scalar quarks to QED,
\begin{align}
\#_{2,2}=\frac{1}{2!}\frac{1}{6}\frac{2N_fN_c}{(4\pi)^2}.
\end{align}
Here, the $(4\pi)^{-2}$ part of the normalisation, already present in \eqref{eq:lag}, arises from taking the trace in \eqref{eq:scalar}, $\frac{1}{2!}$ is due to the Taylor expansion of the Wilson line to the second order in the source $V$, and $\frac{1}{6}$ is the result of carrying out the $[\D y]$ path integral as well as the $\D\hat\tau_j$ integrations after expanding the sources to second order in four-gradients. For details see \eqref{eq:2ndsource} to \eqref{eq:highfree} for $L=1$ in the following subsection. As explained in the last section, the independence from the unphysical value of the regulator $\varepsilon$ can be achieved by reconstructing the full five-dimensional expression
\begin{align}
\Z=\#_{2,2}\iint_\varepsilon^\infty\D^5x\sqrt{g}g^{MK}g^{NL}\V_{MN}\V_{KL},
\label{eq:fivetwo}
\end{align}
where capital indices are summed over all five dimensions. 
The corresponding classical equations of motion read
\begin{align}
g^{NL}\nabla_N\breve\V_{KL}(x_0,T)=0.
\end{align}
Imposing the axial gauge $\breve\V_T=0$ automatically implies Lorenz gauge $\partial\cdot\breve\V=0$.
For the transverse components this means in 4d momentum space
\begin{align}
\Big(\partial_T^2-\frac{p^2}{4T}\Big)\tilde{\breve\V}^\perp(p,T)=0.
\end{align}
The Fourier transformed boundary condition \eqref{eq:inicond},
\begin{align}
\tilde{\breve\V}_\mu(p,T=\varepsilon)=\tilde V_\mu(p),
\end{align}
identifies $p$ with the four-momentum of the source.
The normalisable solution involves Bessel's $K$ (see 9.6.1.~et seq.~in \cite{as}),
\begin{align}
\tilde{\breve\V}^\perp=\tilde V^\perp(p)\sqrt{p^2T}K_1(\sqrt{p^2T}),
\end{align}
for which according to 9.6.28 in \cite{as}
\begin{align}
\partial_T\tilde{\breve\V}^\perp=\tilde V^\perp(p)p^2K_0(\sqrt{p^2T})/2.
\end{align}

Next, we have to put this solution back into the action \eqref{eq:fivetwo} after Fourier transforming it (or, alternatively, we have to transform the solution). Being purely quadratic, $\Z$ on the saddle point amounts to a surface term,
\begin{align}
\breve\Z
&=
4\#_{2,2}\int\D^4x_0\,\eta^{\nu\lambda}[\breve\V_\nu^\perp\partial_T\breve\V_\lambda^\perp]_\varepsilon^\infty
=\\&=
4\#_{2,2}\int\frac{\D^4p}{(2\pi)^4}\,\eta^{\nu\lambda}[\tilde{\breve\V}_\nu^\perp\partial_T\tilde{\breve\V}_\lambda^{\perp*}]_\varepsilon^\infty,
\end{align}
where $\perp$ indicates that only 4d transverse components contribute, and $^*$ stands for the complex conjugate.
Consequently, using 9.6.13.~from \cite{as},
\begin{align}
\breve\Z
&=
-2\#_{2,2}\int\frac{\D^4p}{(2\pi)^4}\eta^{\nu\lambda}\tilde V_\nu^\perp\tilde V_\lambda^{\perp*}p^2K_0(\sqrt{p^2\varepsilon})
=\\&=
\#_{2,2}\int\frac{\D^4p}{(2\pi)^4}\underbrace{\eta^{\nu\lambda}\tilde V_\nu^\perp\tilde V_\lambda^{\perp*}p^2}_{\hat=|\tilde V_{\mu\nu}|^2/2}\{\ln(p^2\varepsilon)+O[(p^2\varepsilon)^0]\}.
\label{eq:div}
\end{align}
Comparing the UV-divergent contributions to the prefactor of the kinetic term $-(4e^2)^{-1}$ (in our conventions $e$ is contained in $V$),
\begin{align}
\frac{1}{4}\beta_1\ln(p^2\varepsilon)\overset{!}{=}\frac{1}{2}\#_{2,2}\ln(p^2\varepsilon)
~~~\Rightarrow~~~
\beta_1=\frac{N_fN_c}{48\pi^2},
\end{align}
which is the known $\beta$-function coefficient for the normalisation adopted here,
\be
\beta_1=\frac{1}{e^3}\frac{\D e}{\D\ln\mu}=-\frac{\D e^{-2}}{\D\ln\mu^2}.
\ee

The computation for fermionic elementary matter proceeds in strict analogy yielding a value for $\#_{2,2}$ that differs by a factor of 4 thus reproducing the corresponding fermionic contribution to the $\beta$-function coefficient\footnote{For the treatment of (non-holographic) renormalisation in the framework of the worldline formalism see \cite{Ritus:1975}.}.

We never used that $\varepsilon$ be small. [In \eqref{eq:div}, we merely presented the behaviour of $\breve\Z$ for if it were small.] Originally, $\varepsilon$ was introduced to regularise the UV divergence of $Z$. Hence, at that point, we had in mind to remove the regulator at the end of the computation by sending it to zero in a controlled manner. At finite $\varepsilon$ the renormalisation condition \eqref{eq:renorm} ascribes the meaning of a scale to $\varepsilon$. If we wanted to keep $\varepsilon$ in its original role, we can introduce counter terms for the divergent piece(s). For \eqref{eq:div}, for example,
\begin{align}
\breve\Z
={}&
\#_{2,2}\int\frac{\D^4p}{(2\pi)^4}\eta^{\nu\lambda}\tilde V_\nu^\perp\tilde V_\lambda^{\perp*}p^2
\times\\&\hskip 2cm\times
\{\ln(\mu^2\varepsilon)+\ln(p^2/\mu^2)+O[(p^2\varepsilon)^0]\},
\nonumber
\end{align}
the first addend inside the braces, which diverges when $\varepsilon\rightarrow0$, must be cancelled by the introduction of a counter term with opposite sign into which finite parts can be included as well. The remaining $\varepsilon$ independent part depends on the scale $\mu^2$ instead, thereby separating the meaning of regulator and scale.

\subsubsection{Higher spin sources/fields\label{sec:hs}}

Above, we presented the special case of a rank-1 source $V$, but sources of any rank contribute to $Z_\varepsilon$. Here we demonstrate that worldline holography also readily applies to them. More generally, the worldline coupling of a rank-$L$ source $W_{\mu_1\dots\mu_L}$ symmetric in all indices, ${W}_{\mu_1\dots\mu_L}={W}_{(\mu_1\dots\mu_L)}$; traceless, $\eta^{\mu_1\mu_2}W_{\mu_1\dots\mu_L}=0$; and transverse, $\partial_\nu\eta^{\nu\mu_1}W_{\mu_1\dots\mu_L}=0$ is given by \cite{Dietrich:2015oba}
\be
\Lag
=
\frac{2\mathcal{N}}{(4\pi)^2}
\int_\mathrm{P}[\D y]\;\E^{-\int_0^T\D\tau[\frac{\dot y^2}{4}-(-\I \dot y \cdot)^L W(x_0+y)/L!]} ,
\label{eq:lagw}
\ee
where $(\dot y \cdot)^L W$ stands for the $L$-fold contraction of $W$ with $\dot y$. After expanding in powers of the sources, gradient expanding the sources, and carrying out the $[\D y]$ as well as $\D\tau$ integrations we have
\begin{align}
Z_\varepsilon
=
\iint_\varepsilon^\infty&\D^5x\sqrt{g}\sum_{n_\partial,n_W}\#_{n_\partial,n_W} 
\times\\\nonumber&\times 
(g^{\four\four})^\frac{n_\partial+L\,n_W}{2}(\partial_\four)^{n_\partial}[W_{\{\four\}}(x_0)T^\frac{1-L}{2}]^{n_W},
\end{align}
where $\{\four\}$ ($\{\five\}$) indicates that all indices that are not mentioned take values in four (five) dimensions; there are only contributions for $n_\partial+L\,n_W$ even.
Accordingly, the corresponding $\varepsilon$ independent five-dimensionally completed formulation, is given by
\begin{align}\label{eq:zw}
\Z
=
\iint_\varepsilon^\infty&\D^5x\sqrt{g}\sum_{n_\partial,n_W}\#_{n_\partial,n_W}
\times\\\nonumber&\times 
(g^{\five\five})^\frac{n_\partial+L\,n_W}{2}(\nabla_\five)^{n_\partial}[\W_{\{\five\}}(x_0,T)]^{n_W},
\end{align}
where $T^{1-L}W(x_0)\rightarrow\W(x_0,T)$. 

For the sake of concreteness, let us determine the coefficients for the terms up to the second order in fields and gradients in \eqref{eq:lagw}. Up to the second order in the fields,
\begin{align}
\Lag
\supset
\frac{(-\I)^{2L}}{2(L!)^2}\frac{2\mathcal{N}}{(4\pi)^2}
\int_0^T\D\tau_1\D\tau_2\int_\mathrm{P}[\D y]\;\E^{-\int_0^T\D\tau\frac{\dot y^2}{4}}
\nonumber\times\\\times
(\dot y_1\cdot)^LW_1(\dot y_2\cdot)^LW_2.
\label{eq:2ndsource}
\end{align}
where $y_j=y(\tau_j)$. Expanding additionally up to the second order in four-gradients,
\begin{equation}
W(x_0+y)=\E^{y\cdot\partial_{x_0}}W(x_0)\approx[1+y\cdot\partial_{x_0}+\sfrac{1}{2}(y\cdot\partial_{x_0})^2]W(x_0),
\nonumber
\end{equation} 
yields
\begin{align}
\Lag
\supset{}&
\frac{(-\I)^{2L}}{2(L!)^2}\frac{2\mathcal{N}}{(4\pi)^2}
\int_0^T\D\tau_1\D\tau_2\int_\mathrm{P}[\D y]\;\E^{-\int_0^T\D\tau\frac{\dot y^2}{4}}
\nonumber\times\\&\times
\{W_0(\cdot\dot y_1)^LW_0(\cdot\dot y_2)^L
+\\&~+\nonumber
[y_1\cdot\partial_{x_0}W_0(\cdot\dot y_1)^L][y_2\cdot\partial_{x_0}W_0(\cdot\dot y_2)^L]
\},
\end{align}
where $W_0=W(x_0)$.
The first order in the gradients and terms where both gradients act on the same source integrate to zero (also taking into account the tracelessness of $W$) and are omitted right away. Performing the $[\D y]$ integration yields 
\begin{align}
\Lag
\supset{}&
\frac{(-\I)^{2L}}{2L!}\frac{2}{(4\pi)^2}
\int_0^T\D\tau_1\D\tau_2
[\ddot P_{12}^L W(\eta^{\four\four})^LW
\nonumber-\\&-
\ddot P_{12}^LP_{12}\eta^{\mu\nu}(\partial_\mu W)(\eta^{\four\four})^L(\partial_\nu W)
\nonumber-\\&-
L\ddot P_{12}^{L-1}\dot P_{12}^2\eta^{\mu\lambda}\eta^{\nu\kappa}(\partial_\mu W_\kappa)(\eta^{\four\four})^{L-1}(\partial_\nu W_\lambda)
],
\end{align}
where, henceforth, we suppress the index 0 to counteract the accumulation of indices. `$(\eta^{\four\four})^L$' represents the $L$-fold contraction of the $2L$ indices of the $W$-s that are not shown explicitly with the inverse flat metric. The worldline propagator $P(\tau_1,\tau_2)\equiv P_{12}$, in the centre-of-mass convention $\int_0^T\D\tau\, y=0$, where it is manifestly proper-time translationally invariant, and its first two derivatives with respect to its first argument read \cite{Strassler:1992zr}
\begin{align}
P_{12}&=|\tau_1-\tau_2|-(\tau_1-\tau_2)^2/T,\\
\dot P_{12}&=\mathrm{sign}(\tau_1-\tau_2)-2(\tau_1-\tau_2)/T,\\
\ddot P_{12}&=2\delta(\tau_1-\tau_2)-2/T.
\end{align}
(The counter charge $-2/T$ on the right-hand side of the last line is required to invert the (one-dimensional) Lapalcian $\partial_\tau^2$ on the compact interval $[0;T]$ and is consistent with the centre-of-mass convention. See also the derivation in \cite{Dietrich:2014ala}.)
Performing the $\D\tau_j$ integrations leads to
\begin{align}
\Lag
\supset{}&
-\frac{2^{L-1}T^{3-L}}{6L!}\frac{2}{(4\pi)^2}
\times\\&\times
(\eta^{\mu\nu}\eta^{\kappa\lambda}-L\eta^{\mu\lambda}\eta^{\nu\kappa})
(\partial_\mu W_\kappa)(\eta^{\four\four})^{L-1}(\partial_\nu W_\lambda).\nonumber
\end{align}
Terms containing $P_{12}$ or $\dot P_{12}$ at coincident proper times $\tau_1=\tau_2$ do not contribute, as $P_{11}=0=\dot P_{11}$. We regularise powers of $\delta$ distributions according to $[\delta(\tau_1-\tau_2)]^l\rightarrow\delta(\tau_1-\tau_2)/T^{l-1}$ or, equivalently, $\ddot P_{12}^L\rightarrow(-2/T)^{L-1}\ddot P_{12}$. The thus-obtained Lagrangian 
\begin{align}\label{eq:highfree}
\Lag
\supset{}
-\frac{2^{L-1}}{6L!}&\frac{2}{(4\pi)^2}
(g^{\mu\nu}g^{\kappa\lambda}-Lg^{\mu\lambda}g^{\nu\kappa})
\times\\&~~~\times
(\partial_\mu W_{\kappa} T^{1-L})(g^{\four\four})^{L-1}(\partial_\nu W_{\lambda} T^{1-L}),
\nonumber\end{align}
pertains to a field theory over AdS$_5$ with all fifth polarisations and gradients zero.
Next, we achieve independence from $\varepsilon$ by reconstructing the complete five-dimensional and locally invariant theory according to the above rules,
\begin{align}
\Z\supset{}
-\frac{2^{L-1}}{6L!}\frac{2}{(4\pi)^2}&\iint_\varepsilon^\infty\D^5x\sqrt{g}
\times\\&\times
(g^{MN}g^{KJ}-Lg^{MJ}g^{NK})
\nonumber\times\\&\times
(\nabla_M\W_K)(g^{\five\five})^{L-1}(\nabla_N\W_J).
\nonumber\end{align}
Expressed with partial instead of covariant derivatives, this amounts to
\begin{align}\label{eq:partder}
\Z\supset{}
\iint_\varepsilon^\infty \D^5x\sqrt{g}
\{&
(g^{MN}g^{KJ}-Lg^{MJ}g^{NK})
\times\\\nonumber&~\times
(\partial_M\W_K)(g^{\bullet\bullet})^{L-1}(\partial_N\W_J)
+\\\nonumber&+
4(L-1)\W(g^{\bullet\bullet})^{L}\W
\},
\end{align}
where the total contribution from the Christoffel symbols amounts to the term without derivatives.
This coincides with the result obtained in \cite{Dietrich:2015oba}.
Varying this effective action with respect to the four-dimensional components, imposing axial gauge, transversality\footnote{Imposing axial gauge and transversality corresponds to adopting the analogue of five-dimensional radiation gauge.}, and tracelessness yields the corresponding components of an AdS Fr{\o}nsdal equation,
\begin{align}
\Big[-T^{1-L}\partial_TT^{L-1}\partial_T-\frac{1}{4}\frac{\Box}{T}
+\frac{L-1}{T^2}\Big]\W_\perp=0.
\label{eq:holocleom}
\end{align}
Analysing the small-$T$ behaviour by means of a power-law ansatz $\W_\perp\propto T^\alpha$ yields the characteristic equation
\be
-\alpha(\alpha-2+L)+L-1=0,
\ee
which is solved by
\be
\alpha=1~~~\&~~~\alpha=1-L.
\label{eq:expo}
\ee
For scalar elementary matter this coincides with the result in \cite{deTeramond:2008ht,Karch:2006pv}.

The equations obtained by varying with respect to fields with at least one $T$ component are given by
\begin{align}
g^{MN}(\nabla_{M}\nabla_{N}\W_{M_1\dots M_L}
-
L\nabla_{M}\nabla_{(M_1}\W_{M_2\dots M_L)N})
=0,\nonumber
\end{align}
where we have imposed the axial gauge only a posteriori and did {\sl not} insist on transversality. Equations with more than two $T$ components vanish identically after imposing the gauge condition. The equation with exactly two $T$ components amounts to 
$
\frac{\partial\ln\mathrm{tr}\W}{\partial\ln T}=\mathrm{const.};
$
the equation with exactly one $T$ component states that a linear combination of $\mathrm{div}\,\W$, $\frac{\partial\mathrm{div}\W}{\partial\ln T}$, and $\mathrm{grad}\,\mathrm{tr}\,\W$ is zero.\footnote{Technically, this is due to the fact that only the Christoffel symbols with an odd number of components in the $T$ direction are nonzero for \eqref{eq:patch}: $${\Gamma^T}_{\mu\nu}\overset{g}{=}2\eta_{\mu\nu},~~~
{\Gamma^\mu}_{T\nu}\overset{g}{=}-\sfrac{1}{2T}\delta^\mu_\nu={\Gamma^\mu}_{\nu T},~~~
{\Gamma^T}_{TT}\overset{g}{=}-\sfrac{1}{T}.
$$} Consequently, if $\W$ is transverse as well as traceless on the boundary, it will remain so everywhere in the bulk\footnote{If we had enforced transversality from the very beginning of the derivation of the effective action (We only used the tracelessness.), the ($L$ dependent) cross term would be absent.
Then the fifth components of the saddle-point condition would be force the trace and the non-transverse components to vanish identically, a fact that had been observed before in \cite{deTeramond:2013it}.}. Thus, the variation with respect to the fifth-dimensional polarisations (together with the boundary conditions) enforces transversality and tracelessness\footnote{There are several approaches to obtaining the Fr{\o}nsdal equations for transverse and traceless fields from an un- or less constrained variational principle, like auxiliary compensator fields \cite{FierzPauli,Singh:1974qz} and the related relaxation to double tracelessness \cite{Fronsdal:1978rb}.}.

For the scalar source $\W_{L=0}$, \eqref{eq:partder} features a tachyonic mass term, which saturates the Breitenlohner-Freedman bound. For the boundary condition $\lim_{T\rightarrow 0}(\W_{L=0} T^{L-1=-1})=m^2$, the equation of motion \eqref{eq:holocleom} is solved by $\W_{L=0}= m^2T$. This is the tachyon (squared) profile \cite{Bigazzi:2005md} for a free theory of elementary matter with the explicit mass $m$. 

Taking stock, the present framework links a free scalar quantum field theory on Mink$_4$ with sources of any spin to a field theory for these sources on AdS$_5$. Such a duality was conjectured to exist \cite{Sundborg:2000wp}.

\subsubsection{Spin-2 revisited\label{sec:revisited}}

For rank-2 sources the explicit results above corresponds to the linearised Einstein equations. In this context, the rank-2 source represented the deviation $h_{\mu\nu}$ from the Minkowski metric $\eta_{\mu\nu}$. A straightforward expansion to finite powers of the deviation, however, does not posses full diffeomorphism invariance, but, like for the vector case in footnote \ref{foot1}, it is possible to devise a fully covariant expansion scheme. In short, while the derivation in the vector case makes use of the Fock-Schwinger gauge to express the vector field in terms of   covariant derivatives (and their commutator, the field tensor), the spin-2 case makes use of Riemann normal coordinates, where the full metric $\g_{\mu\nu}=\eta_{\mu\nu}+h_{\mu\nu}$ is expressed in terms of curvature tensors constructed from the corresponding Levi-Civita connection $\nabla[\g]$. The expansion again takes the form
\begin{align}\nonumber
Z_\varepsilon
&=
\int_\varepsilon^\infty\frac{\D T}{2T^3}\int \D^4x_0\sqrt{\g}\sum_n \#_n (T\g^{\four\four})^n(\nabla_\four[\g])^{2n}
=\\&=
\iint_\varepsilon^\infty\D^5x\sqrt{\fg}\sum_n \sharp_n (\fg^{\four\four})^n(\nabla_\four[\g])^{2n},
\label{eq:dewitt}
\end{align}
where $\fg$ stands for the five-dimensional Fefferman-Graham \cite{Feffermann:1985} embedding of $\g$,
\begin{align}
\D s^2\overset{\fg}{=}\natural\Big(\frac{\D T^2}{4T^2}+\frac{\g_{\mu\nu}\D x^\mu\D x^\nu}{T}\Big),
\end{align}
and $\sharp_n=\#_n\natural^{n-5/2}$. Again, \eqref{eq:dewitt} is not a differential operator. There are no partial derivatives acting to the right. The expansion is to symbolise all occurring combinations; the commutator of two covariant derivatives, for example, yields the Riemann tensor.

The independence \eqref{eq:renorm} from $\varepsilon$ can once more be achieved by the five-dimensional completion 
\begin{align}
\Z
=
\iint_\varepsilon^\infty\D^5x\sqrt{\fg}\sum_n \sharp_n (\fg^{\five\five})^n(\nabla_\five[\fg])^{2n}
\end{align}
and subsequent evaluation of the action on the saddle point with the boundary condition 
\begin{align}
\breve\fg_{\mu\nu}(x_0,T=\varepsilon)=\frac{\natural}{\varepsilon}\g_{\mu\nu}(x_0)
\end{align}
and the gauge condition
\begin{align}
\breve\fg_{TN}\overset{!}{=}\natural g_{TN}\,\forall N,
\end{align}
with $g$ from \eqref{eq:patch}, which corresponds to the absence of deviations with fifth-dimensional polarisations, 
\begin{align}
h_{TN}\overset{!}{=}0\,\forall N.
\end{align}

Let us study the leading terms. The $\#_n$ are the DeWitt-Gilkey-Seeley coefficients \cite{DeWitt:1965}. The first two correspond to a negative cosmological constant and an Einstein-Hilbert term,
\begin{align}
\Z\supset\frac{1}{6(4\pi)^2}\iint_\varepsilon^\infty\D^5x\sqrt{\fg}(R[\fg]+6).
\label{eq:eh}
\end{align} 
As a consequence, the corresponding Einstein equations admit an AdS$_5$ solution with the squared AdS curvature radius 
\begin{align}
\sharp=R_\mathrm{AdS}^2=\frac{(5-1)(5-2)}{6}=2.
\end{align}
Taking into account the boundary and gauge conditions the solution reads
\begin{align}
\D s^2\overset{\breve\fg}{=}2\Big(\frac{\D T^2}{4T^2}+\frac{\D x_0\cdot\D x_0}{T}\Big).
\label{eq:fgsad}
\end{align}
Consequently, at least to this order, an AdS background is self-consistently maintained by the formalism. (To higher orders, a space of constant curvature remains a saddle-point solution albeit with a different curvature radius.)
The isometries of any AdS$_5$ space, i.e., with any value of the curvature radius, coincide with the conformal group over Mink$_4$ (just like Mink$_4$ is Poincar\'e invariant for any value of the speed of light). Therefore, the value of the AdS radius is of secondary importance insofar as it does not alter the structure of the present result. For example,
\begin{align}
\Z
=
\iint_\varepsilon^\infty\D^5x\,\breve\fg^{1/2}&\sum_{n_\partial,n_V}\sharp_{n_\partial,n_V}
\times\\&\times
 (\breve\fg^{\five\five})^\frac{n_\partial+n_V}{2}(\nabla_\five[\breve\fg])^{n_\partial}[\V_\five(x_0,T)]^{n_V},
\nonumber
\end{align}
where $\sharp_{n_\partial,n_V}=\#_{n_\partial,n_V}\natural^{(n_\partial+n_V-5)/2}$, is an identical reexpression for \eqref{eq:calz}, which is independent of $\natural$. Also the covariant derivatives do not depend on the curvature radius. [Consistently, neither does the (1,3) Riemann tensor.] Likewise,
\begin{align}
\Z
=
\iint_\varepsilon^\infty&\D^5x\,\breve\fg^{1/2}\sum_{n_\partial,n_W}\sharp_{n_\partial,n_W}
\times\\\nonumber&\times 
(\breve\fg^{\five\five})^\frac{n_\partial+L\,n_W}{2}(\nabla_\five[\breve\fg])^{n_\partial}[\W_{\{\five\}}(x_0,T)]^{n_W}
\end{align}
where $\sharp_{n_\partial,n_W}=\#_{n_\partial,n_W}\natural^{(n_\partial+L\,n_W-5)/2}$ is identical to \eqref{eq:zw} and independent of $\natural$.

\subsection{Infrared\label{sec:ir}}

The worldline holographic framework can also handle variations of infrared scales. Consider
\begin{align}
Z_{\varepsilon,k^2}
=&
\iint_\varepsilon^\infty\D^5x\sqrt{g}
\,\rho(k^2T)
\sum_{n_\partial,n_V} \#_{n_\partial,n_V}
\times\\&\hskip 2.5cm\times\nonumber
 (g^{\four\four})^\frac{n_\partial+n_V}{2}(\partial_\four)^{n_\partial}[V_\four(x_0)]^{n_V},
\end{align} 
where $\rho(k^2T)\xrightarrow{k^2T\rightarrow\infty} 0$ and $\rho(k^2T)\xrightarrow{k^2T\rightarrow0} 1$ \cite{Oleszczuk:1994st}. This can, for example, be realised by a sharp proper-time cutoff $\rho(k^2T)=\theta(k^2T-1)$ or by a mass term \cite{Strassler:1992zr} $\rho(k^2T)=\exp(-k^2T)$. We could now analyse the fate of the scale $k^2$ by repeating the steps from the beginning of Sect.~\ref{sec:fifth}. From Sect.~\ref{sec:ads}, however, we already know that worldline holography ensures the independence from an overall scale through the isometries of AdS$_5$ acting on,
\begin{align}
\Z_{\varepsilon, k^2}
&=
\iint_\varepsilon^\infty\D^5x\sqrt{g}\,\rho(k^2T)\sum_{n_\partial,n_V}\#_{n_\partial,n_V}
\times\\\nonumber&\hskip 2cm\times
 (g^{\five\five})^\frac{n_\partial+n_V}{2}(\nabla_\five)^{n_\partial}[\V_\five(x_0,T)]^{n_V}.
\end{align} 
Since we are thus able to eliminate the dependence on such an overall scale, the final result can only depend on the combination $k^2\varepsilon$. This corresponds to the renormalisation condition $\varepsilon\partial_\varepsilon Z_{\varepsilon, k^2}=k^2\partial_{k^2}Z_{\varepsilon, k^2}$. This is the requirement that the effective action do not change under simultaneous changes of $\varepsilon$ and $k^2$ such that $\varepsilon k^2$ remains the same. The renormalisation condition is solved by the saddle point expression
\begin{align}
\breve\Z_{\varepsilon k^2}
&=
\iint_\varepsilon^\infty\D^5x\sqrt{g}\,\rho(k^2T)\sum_{n_\partial,n_V}\#_{n_\partial,n_V}
\times\\\nonumber&\hskip 2cm\times
 (g^{\five\five})^\frac{n_\partial+n_V}{2}(\nabla_\five)^{n_\partial}[\breve\V_\four(x_0,T)]^{n_V}.
\end{align} 
Accordingly, after an introduction of counter terms like in \ref{sec:qed} and taking the limit $\varepsilon\rightarrow0$ subsequently, the resulting effective action will only depend on the combination $k^2/\mu^2$.
In a UV finite theory, and more generally in all the UV finite terms, $\varepsilon$ is not needed in its role as UV regulator, and we can consider the case where it is zero. Then the alternative renormalisation condition $k^2\partial_{k^2}\ln Z_{0,k^2}\overset{!}{=}0$ is solved by $\breve\Z_{0, k^2}$. 
We continue the discussion of infrared scales elsewhere.

\section{Short summary\label{sec:sum}}

Worldline holography maps a $d$ dimensional quantum field theory onto a $d+1$ dimensional field theory for the sources of the former, to all orders in the elementary fields and sources of any rank. The metric of the $d+1$ dimensional space is obtained by a Fefferman-Graham embedding \cite{Feffermann:1985} of the $d$ dimensional one. For Mink$_d$ this gives the known AdS$_5$. Above, we have shown that worldline holography is the solution to a Wilson-Polchinski renormalisation condition \eqref{eq:renorm}, which ensures the independence of physical quantities from the ultraviolet regulator. [Infrared scales can be treated analogously. (See Sect.~\ref{sec:ir}.)] Said renormalisation condition serves as (part of) the definition of the worldline holographic framework in general and of the fifth-dimensional profiles of the sources in particular. For the cases studied here the result is exactly the same as the one \cite{Dietrich:2013kza,Dietrich:2015oba} obtained by optimising a Wilson (gradient) flow \cite{Luscher:2009eq}. As crosschecks we holographically reconstructed the leading QED beta-function coefficient in Sect.~\ref{sec:qed}. In Sect.~\ref{sec:hs}, we derived the worldline holographic dual for a free scalar field theory on Mink$_4$: a field theory for sources turned fields of all integer spins over AdS$_5$, which was postulated in \cite{Sundborg:2000wp}. A manifestly diffeomorphism invariant expansion of the rank-2 case leads to the DeWitt-Gilkey-Seeley coefficients \cite{DeWitt:1965}. In Sect.~\ref{sec:revisited}, this serves to show that AdS$_5$ is a selfconsistent solution of the worldline holographic framework.

\section{Further discussion and outlook\label{sec:out}}

Barring anomalies, a quantised version of $\ln\Z$ would bear the necessary isometries to be a solution of the renormalisation condition \eqref{eq:renorm} once the appropriate boundary conditions \eqref{eq:inicond} are imposed. The saddle point remains the leading contribution. The first correction is given by the fluctuation determinant.\footnote{The use of the worldline formalism allows us to make a link \cite{Dietrich:2007vw} to the Gutzwiller trace formula \cite{Gutzwiller:1971fy}, which describes quantum systems through classical attributes as well.} In some situations, like the case discussed in Sect.~\ref{sec:hs}, however, the distinction between a `quantum' and a `classical' answer ultimately turns out to be irrelevant, as the quantum contributions from the individual fields of different rank over AdS cancel when summed over the complete tower of higher-spin states \cite{Giombi:2013fka}. 

Furthermore, we are free to again carry out the quantum computation in the worldline formalism. In the course of this computation $g$ (or $\breve\fg$) is Fefferman-Graham embedded into 
\begin{align}
\D s^2=\frac{\D\Theta^2}{4\Theta^2}+\frac{g_{MN}\D x^M\D x^N}{\Theta},
\end{align}
where $\Theta$ is the new proper time. The isometries of the five-dimensional part are preserved, but, while still being conformally flat, this six-dimensional space does not have constant curvature and no enhanced scaling symmetry. Rather, $\Theta$ dials through the curvature radius of the five-dimensional part. Additionally, the six-dimensional effective action depends on functions of $\natural\Theta$, which consistently forestall a higher scaling symmetry.

~\\

The latter is only one more example for starting worldline holography from a spacetime other than Mink$_4$, but the pattern already shines through: The lower-dimensional metric will be embedded into a higher-dimensional Fefferman-Graham metric, and additional curvature (of the original space) dependent terms arise in the effective action.

Above, we were predominantly investigating worldline holography linking a quantum field theory over four-dimensional Minkowski space to a field theory for its sources over five-dimensional anti-de Sitter space, more precisely AdS$_{4,1}$. Then again, for the worldline approach we first Wick rotated to Euclidean space and from there found a connection to five-dimensional hyperbolic space H$_5$, i.e., AdS$_{5,0}$. Analytically continuing the time direction afterwards leads to AdS$_{4,1}$. This amounts to changing $\epsilon_t$ from $-1$ to $+1$ in
\be
\D s^2\overset{g}{=}\epsilon_T\frac{\D T^2}{4T^2}+\frac{\epsilon_t(\D t)^2+|\D \vec x|^2}{T},
\ee
for $\epsilon_T=+1$. Another analytic continuation \cite{Maldacena:2002vr} taking $\epsilon_T$ from $+1$ to $-1$ links five-dimensional de Sitter space dS$_5$ with H$_5$ or AdS$_{2,3}$ with AdS$_{4,1}$,
\be
\nonumber
\xymatrix{
\mathrm{dS}_5 \ar@{<->}[r]^{+\epsilon_t-} \ar@{<->}[d]_{\underset{+}{\overset{-}{\epsilon_T}}} & \mathrm{AdS}_{2,3}   \ar@{<->}[d]\ar@{<->}[d]^{\underset{+}{\overset{-}{\epsilon_T}}} \\ 
\mathrm{H}_5  \ar@{<->}[r]_{+\epsilon_t-}              & \mathrm{AdS}_{4,1}
}
\ee
All these are holographic pictures of four-dimensional flat spacetimes.

Moreover, in the imaginary-time formalism for thermal field theory the time $t$ is compactified with the period of the inverse temperature. The straightforward application of worldline holography yields thermal AdS space, i.e., AdS space with a compactified temporal direction. There is, however, a second space with the same boundary topology and identical source configurations on the boundary, the AdS black hole \cite{Hawking:1982dh}. Both are stationary points of the action \eqref{eq:eh}. The preferred configuration is selected by the relative value of the action. In the present setting the relative importance of bosonic and fermionic degrees is decisive for which of the two five-dimensional spacetimes is preferred \cite{Dietrich:2015dka}.

Furthermore, worldline holographic duals are also available in the non-relativistic setting \cite{Dietrich:2013kza,Dietrich:2015oba}. Representing the conformal Galilean symmetry of the Schr\"odinger equation by imposing a fixed light-cone momentum $p^+=m$ \cite{Son:2008ye} on 4+1 dimensional Minkowski space, which selects $x^+$ as normal time, in the worldline approach we find a six-dimensional line element
\be
\D^2s\overset{\mathrm{g}}{=}
-\frac{\D T^2}{4T^2}
+\frac{2(\D x^+)^2}{T^2}
+\frac{2\D x^+\D x^--\D\mathbf{x}\cdot\D\mathbf{x}}{T} ,
\ee
with the correct volume element $\sqrt{\mathrm{g}}\propto T^{-7/2}$ thus reproducing \cite{Son:2008ye,Balasubramanian:2008dm}.
(The power of $T$ in the denominator of the $(\D x^+)^2$ term can be different \cite{Balasubramanian:2008dm} without influencing the volume element.)

~\\

So far, we mostly studied two-point functions and kinetic terms. Exceptions are, e.g., the non-Abelian part of the vector kinetic term and covariant expressions for spin-2 backgrounds. Worldline holography also gives a general prescription for determining interactions. 
Any number of terms can be worked out this way.
While this is expected to be possible consistently over an AdS background \cite{Fradkin:1986qy}, there are known obstacles over others like Minkowski or de Sitter. In view of the fact that our prescription can yield various $d+1$ dimensional spacetimes (see below) a thorough study of interactions in our framework, in particular of the higher-spin fields, is an important future task.

~\\

In this paper we have concentrated on scalar elementary matter, in order not to shroud the structure of the worldline holographic framework by carrying along additional degrees of freedom. It is, however, fermionic elementary matter, which is realised in nature. Fermions do not pose any additional fundamental challenges to worldline holography, but have a richer phenomenology---especially also from the vantage point of the present framework---, due to their spin degree of freedom. The corresponding results are presented elsewhere \cite{Master,Fermions}.

~\\

\section*{Acknowledgments}

The authors would like to thank
Stan Brodsky,
Guy de T\'eramond,
Luigi Del Debbio,
Florian Divotgey,
Gerald Dunne,
Gia Dvali,
Joshua Erlich,
J\"urgen Eser,
Francesco Giacosa,
C\'esar G\'omez,
Stefan Hofmann,
Paul Hoyer,
Johannes Kirsch,
Sebastian Konopka,
Matti J\"arvinen,
Yaron Oz,
Stefan Rechenberger,
Dirk Rischke,
Ivo Sachs,
Andreas Sch\"afer,
Stefan Theisen,
and
Roman Zwicky
for discussions.

\end{document}